\documentstyle[11pt,newpasp,twoside,epsf]{article}
\begin{document}
\title{Halo enrichment by disrupted globular clusters}      
\author{Holger Baumgardt}
\affil{Department of Mathematics and Statistics, University of Edinburgh, King's
Buildings, Edinburgh EH9 3JZ, UK} 

\begin{abstract}
We study the evolution of the galactic globular cluster system to determine its
initial mass-function and the
fraction of halo stars that could have come from disrupted globular clusters.
We study the cluster evolution under the influence of two destruction mechanisms: Two-body
relaxation and dynamical friction. New results of $N$-body simulations are used for
the lifetimes of clusters dissolving under the influence of two-body relaxation.  

Two different mass-functions are studied: A gaussian initial distribution
in $\log(M_C)$ with mean mass and scatter similar to what one observes for the galactic 
globular cluster system, and a power-law distribution resembling what is seen for clusters in merging 
and interacting galaxies. We find that in the inner parts of the 
galaxy, both distributions evolve in such a way that they are consistent with the observations.
In the outer parts, a gaussian initial distribution gives the better fit. 
This might change
however if elliptic orbits are considered, or there are undiscovered low-mass clusters 
in the galactic halo. In any case, only a small fraction of 
the stellar halo of the Milky Way originated in globular clusters.  
\end{abstract}

\section{Introduction}

Globular clusters are among the oldest objects in galaxies, and understanding the details
of their formation and evolution can bring valuable insight into the early history of 
galaxies. One of the most interesting questions concerns the relation of globular clusters
to the stellar halos of galaxies. 
At present, only a small fraction of halo stars are found in clusters.
Observations indicate however that the galactic 
halo has a clumpy structure (Majewski et al.\ 1996, Helmi et al.\ 1999), and tidal streams surrounding globular
clusters (Grillmair et al.\ 1995, Leon et al.\ 2000)
have been detected. A significant fraction of the stellar halo  
might therefore have formed in compact stellar systems, although dwarf galaxies are also possible
contributers.

Linked to this question is the determination of the initial mass function of globular clusters.
The globular cluster system of the Milky Way as we observe it today is characterized by a 
gaussian
distribution in absolute magnitudes, with mean $M_V = -7.4$ and scatter $\sigma_M = 1.15$
(Harris 2000). Similar distributions have been found for old cluster systems in other galaxies 
(Kundu et al.\ 1999). In contrast, young massive clusters in interacting and starburst galaxies 
have
power-law distributions over masses $N(M) \sim M^{-\alpha}$, with slopes close to $\alpha = 2$
(Whitmore \& Schweitzer 1995,
Whitmore et al.\ 1999). 
If the globular cluster system of the Milky Way has started with such a mass-function,
it contained many low-mass clusters which dissolve completely and cause 
a stronger enrichment of the stellar halo.

In this paper, we study cluster systems starting with gaussian or power-law mass-functions.
We use analytic formulas for the lifetimes of star clusters and remove clusters with
lifetimes smaller than a Hubble time. 
The number and masses of surviving clusters
are then compared with the Milky Way globular cluster system in order to determine the mass
that must have been initially in the cluster system.
This estimate is then compared with the mass of the stellar halo.
The paper is organised as follows: In section 2 we discuss the dissolution processes of globular
clusters. Section 3 describes the 
set-up of the cluster system and in section 4 we present the results. In section 5 we draw
our conclusions.   

\section{Dissolution processes of star clusters}

Star clusters dissolve due to several processes: In the beginning the loss of gas out of
which the cluster was formed and the mass-loss of individual cluster stars are most important.
Later clusters dissolve mainly due to two-body relaxation and
disc or bulge shocks
from the varying external tidal field. Clusters are also affected by dynamical friction,
which causes them to sink towards the galactic center where they are destroyed by
the strong tidal field. We consider clusters moving in circular orbits
and start
our simulations after the initial gas is removed from the clusters and massive stars
have gone supernova. Hence, we are left with only two dissolution processes: Two-body relaxation
and dynamical friction.

\subsection{Two-body relaxation}

The lifetimes against two-body relaxation are obtained from a fit to the results of $N$-body 
simulations made by Sverre Aarseth and Douglas
Heggie as part of the 'Collaborative Experiment' (Heggie et al.\ 1998). They performed
simulations of multi-mass, King $W_0 = 3.0$ clusters which moved on circular orbits through an 
external tidal field.
Figure 1 shows the
half-mass times of their clusters as a function of the particle number $N$. The best fit is obtained by  
\begin{equation} 
T_{Half} = 0.375 \; \left( \frac{N}{\log(\gamma N)} \right)^{0.8} \; T_{Cross} \; \; ,
\end{equation} 
where $\gamma = 0.02$ (Giersz \& Heggie 1996). The lifetimes are obtained by assuming that they are
twice as large as the half-mass times. We note that the lifetimes of clusters in external tidal
fields increase more slowly with the particle   
number than their relaxation times. It has been shown (Fukushige \& Heggie 2000, Baumgardt 2001)
that this is a consequence of the 
finite escape time of stars from such clusters. As a result,
lifetimes of globular clusters are smaller than previously predicted.

\begin{figure}[t]
\plotone{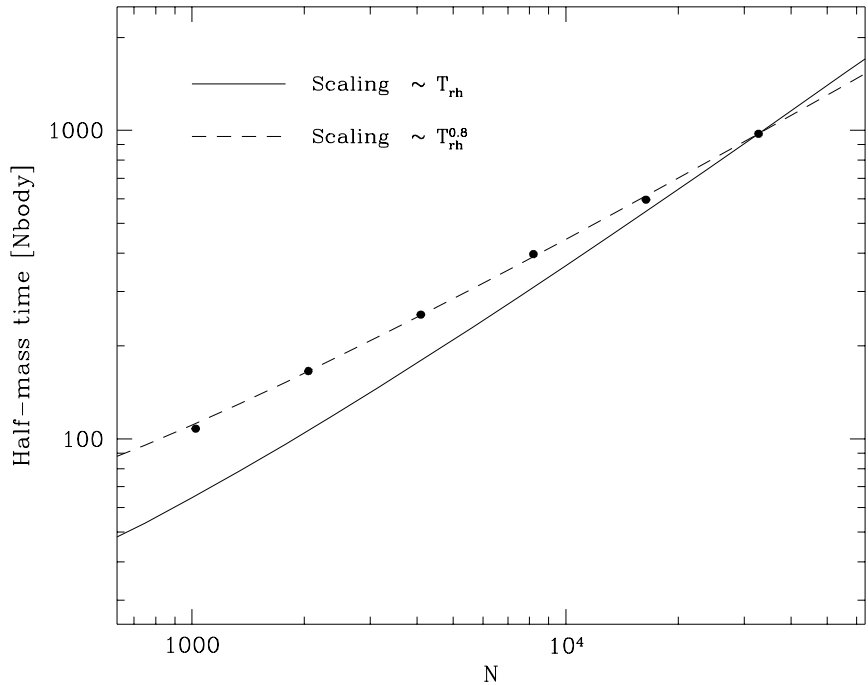}
\caption{Half-mass times as a function of the number of cluster stars for multi-mass    
clusters. The data is taken from the 'Collaborative Experiment' (Heggie et al.\ 1998). The solid line shows a 
scaling proportional to the relaxation time, fitted to the result
of the largest run. It does not fit the data. The dashed line shows the fit for the lifetimes used in
this work. It increases more slowly with the particle number.}
\end{figure}

\subsection{Dynamical friction}

According to Binney \& Tremaine (1987, eq.\ 7-26), the time it takes for a cluster with mass 
$M_C$ and
initial galactocentric distance $R_G$ to sink to the galactic center is given by

\begin{equation}
T_{Fric} = \frac{2.64 \; 10^{11} \; \mbox{yr}}{\ln \Lambda} \; \left( \frac{R_G}{2 \; \mbox{kpc}} \right)^2 
\; \left( \frac{v_{circ}}{250 \; \mbox{km s$^{-1}$}} \right) \; \left( \frac{10^6 \; 
M_{\sun}}{M_C} \right) \;\; .   
\end{equation} 
Here $v_{circ}$ is the circular velocity of the galaxy and
$\ln \Lambda$ is of order 10.

\section{The cluster set-up}

The Milky Way is modeled as an isothermal sphere with circular velocity $v_0 = 220$ km/sec.
We adopt an age of $T_{Hubble} = 12$ Gyr for the age of the galactic globular cluster system.  
Cluster masses either follow a gaussian distribution with mean $\log M_C = 5.15$ and width
$\sigma_M = 0.6$, or a power-law $N(M_C) \sim M_C^{-\alpha}$ with slope $\alpha = 2.0$.
If clusters are distributed according to a power-law, lower and upper limits of $M_C = 10^4 \; M_{\sun}$
and $10^7 \; M_{\sun}$ are chosen for the cluster masses. The lower limit has no influence on the final 
mass-function, since
clusters with $M_C < 10^4 \; M_{\sun}$ dissolve at all galactocentric radii in less than a Hubble time.

We study the evolution
of star clusters between galactocentric distances 1~kpc~$<~R_{GC}~<$~40~kpc. At each $R_{GC}$,
we simulate the evolution of 200.000 clusters. We first calculate their dynamical friction
time $T_{Fric}$ according to eq.\ 2, and remove clusters with $T_{Fric} < T_{Hubble}$. For the 
remaining clusters,
we calculate their tidal radius from the cluster mass and the galactocentric distance.
We assume that the density distribution of a cluster follows a King $W_0 = 3.0$ profile, in which 
case the 
ratio of the half-mass 
radius of a cluster to its tidal radius is given by $r_h/r_t= 0.268$. We can then calculate the 
crossing times $T_{Cross}$ and the dissolution times of the clusters according to eq.\ 1. If the 
dissolution time is smaller than a Hubble time, a cluster 
dissolves completely, otherwise the final mass of the cluster is calculated according to 
\begin{equation} 
M_{Fin} = M_{Ini} \left( 1 - \frac{T_{Hubble}}{T_{Diss}} \right) \;\; ,
\end{equation}  
i.e.\ we assume a constant mass-loss rate.

\section{Results}

Fig.\ 2 shows the evolution of the mass-function of globular clusters at two galactocentric radii. A 
gaussian initial mass-function stays everywhere close to the initial distribution. In the inner
parts, the mean mass increases slightly due to the efficient destruction of low-mass clusters. In the 
outer parts,
the distribution changes very little since only few clusters are destroyed.

A power-law distribution shows a stronger depletion of clusters. In the inner parts, it evolves into
a distribution which is very similar to the one obtained from a gaussian initial mass-function.
The two distributions differ
mainly at large galactocentric radii, where the power-law distribution leads to a mass-function
with lower mean mass. But even at large radii, the power-law distribution evolves into a bell-shaped 
curve.

\begin{figure}[t]
\plotone{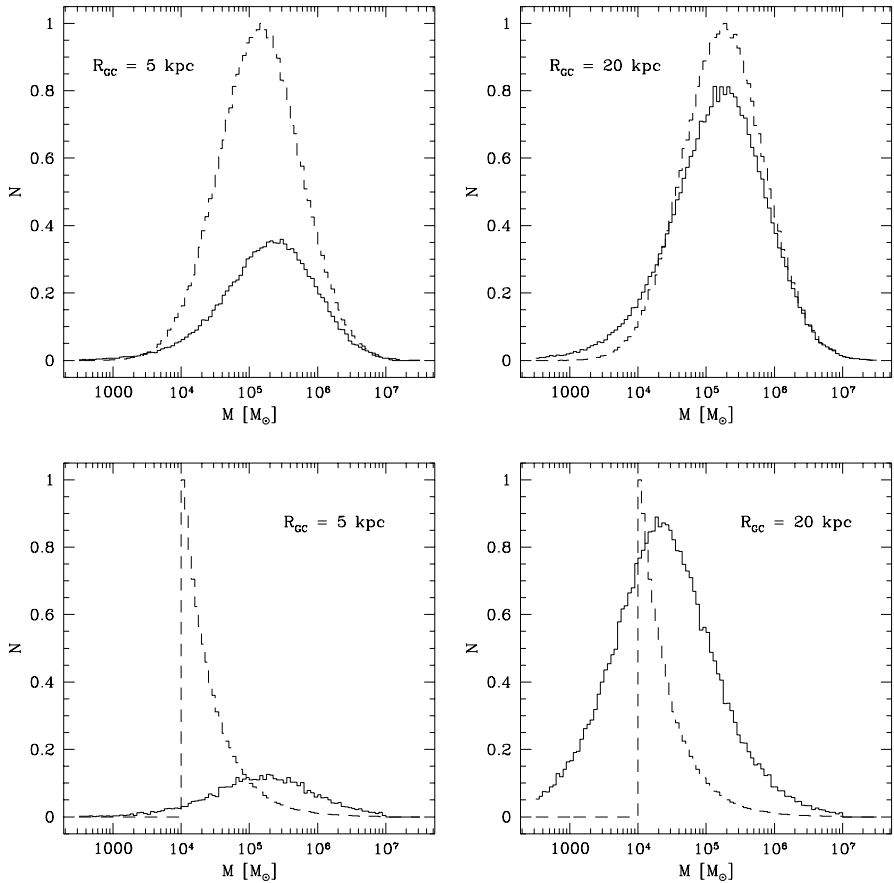}
\caption{Mass functions of globular clusters for two galactocentric distances. The top panels show
the evolution of a gaussian initial mass function, the bottom panels that of a power-law distribution.
The dashed lines show the initial distributions, the solid lines the final ones. The number of final
clusters in the bottom panels are multiplied by a factor of ten to show their distribution.}
\end{figure}

Fig.\ 3 compares the mean cluster masses with the observations of Milky Way clusters.
The properties of galactic globular clusters show no strong variation with galactocentric distance. 
The mean cluster mass might be slighter larger around $R_{GC} = 10$ kpc, but
the differences are small. Independent of the starting condition, the final cluster systems match
the Milky Way clusters for $R_{GC} < 8$ kpc fairly well. Hence, the initial state cannot
be decided from observations of the inner clusters. In the outer parts, a gaussian 
initial distribution
is close to the observations while a power-law mass-function produces a cluster system which has too
many low-mass clusters.
This could change however if elliptic cluster orbits are considered, in which case clusters
are destroyed more easily due to the varying tidal field. In addition,  
undiscovered low-mass clusters could exist in the galactic halo, causing the mean cluster mass
to be lower than observed.

\begin{figure}[t]
\plotone{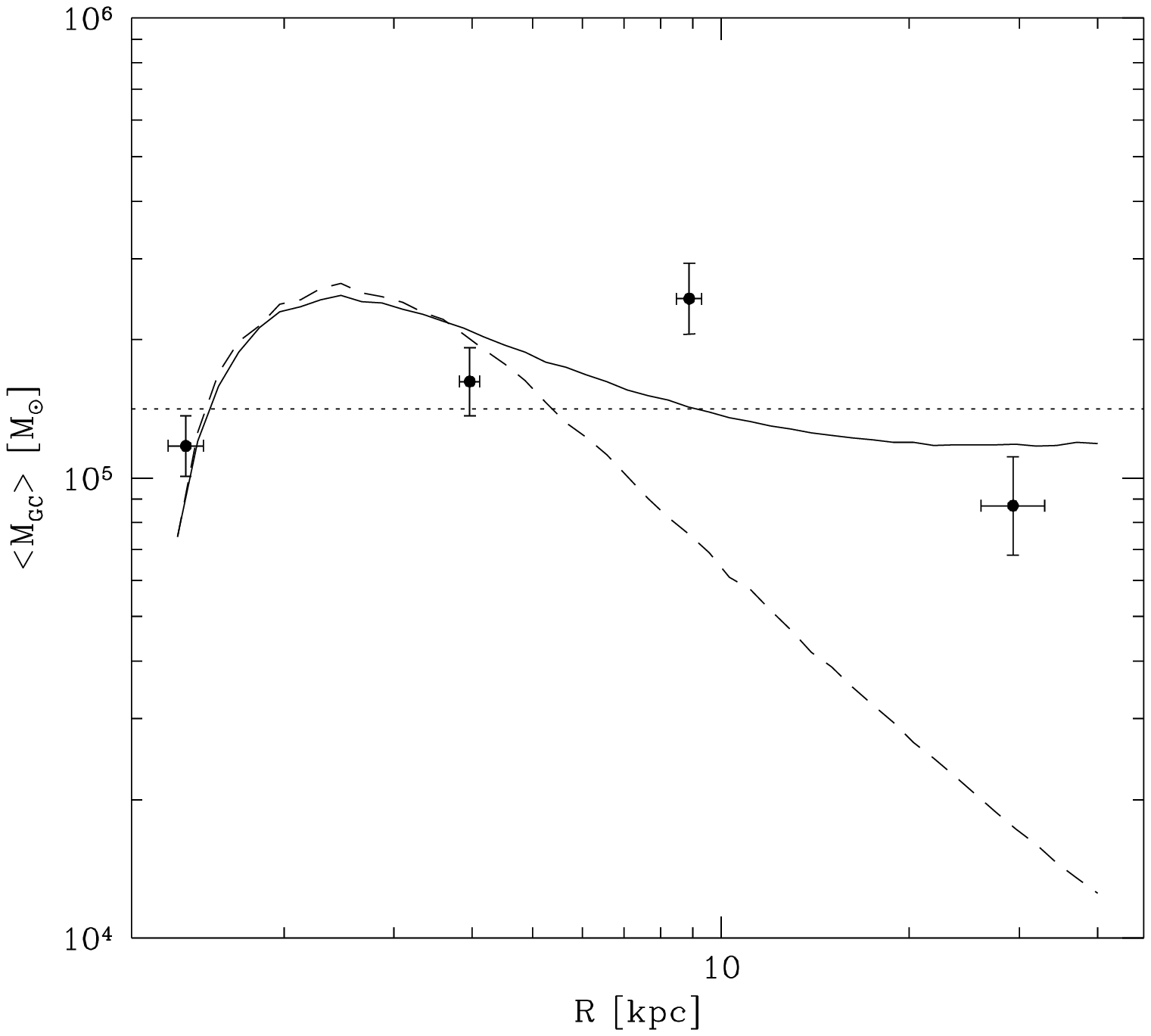}
\caption{Mean masses of globular clusters as a function of galactocentric distance. Mean masses obtained
from a gaussian initial mass function are shown by a solid line, those from a power-law distribution
by a dashed line. The dotted line indicates the mean mass of galactic globular clusters, which is
also the starting mass of the gaussian distribution. Crosses indicate mean masses of galactic globular 
clusters for different galactocentric distances.}
\end{figure}

The fraction of clusters that survive after 12 Gyr and the mass lost from the globular cluster
system depend on the radial distribution of clusters. The number of observed globular clusters
decreases as $\rho \sim R_{GC}^{-3.5}$, at least for $R_{GC} > 4$ kpc (Harris 2000). If we
adopt the same distribution for the initial cluster system, we find that 
for a gaussian mass-function, 
$N_f/N_i = 32~\%$ of globular clusters survive over the whole range of distances studied (1 - 40 kpc).
They contain 36~\% of the mass that was initially in the cluster system. The corresponding
numbers for a power-law distribution are 8 \% and 25 \% respectively. As expected much fewer
clusters survive. However, since most of the dissolving clusters are of low mass, the mass-loss rates 
differ not very much.\footnote{If clusters start with a power-law
mass-function, the fraction of mass surviving in clusters depends on the chosen lower and upper limits.
The dependance is not very strong however. Choosing for example $10^3 \, M_{\sun}$ as a lower limit
instead of $10^4 \, M_{\sun}$, would decrease the mass fraction surviving  
in clusters to 3/4 of its original value.}

According to the McMaster-database of globular cluster parameters (Harris 1996), 
the galactic globular cluster system contains 
132 clusters between 1 and 40 kpc.
These clusters have a total mass of $M_{Clus} = 3.3 \cdot 10^7 \; M_{\sun}$. 
The initial mass in the cluster system was therefore of order 1 to $2 \, \cdot 10^8 \; M_{\sun}$. 
According to Suntzeff et al.\ (1991), the mass of the stellar halo between 4 and 25 kpc is $9 \, \cdot 
10^8 \;
M_{\sun}$. Assuming a $\rho \; \sim \; R_{GC}^{-3.5}$ density law for the halo stars, the halo mass
between 1 and 40 kpc turns out to be $2.6 \, \cdot 10^9 \; M_{\sun}$, which is an order of magnitude more
than the mass of the globular cluster system in this distance range.
Hence, the majority of halo stars was not born in clusters.       

\section{Conclusions}

The evolution of the galactic globular cluster system under the influence of two-body relaxation and
dynamical friction was studied. It was found that a gaussian initial distribution fits
the present day galactic globular cluster system very well. A power-law initial distribution cannot
be ruled out however since it gives a good fit to the observations in the inner parts. The differences
in the outer parts could vanish if additional destruction mechanisms are considered. Furthermore, our 
knowledge of remote halo clusters might not be complete. We cannot constrain the
fraction of stars born in low-mass clusters with $M_C < 10^4 \; M_{\sun}$, since low-mass 
clusters dissolve completely even at the largest distances studied.    
 
Depending on the initial mass-function, only between 5 \% and 10 \%  of the halo stars were born in 
globular clusters. This value is smaller than the fraction 
of stars formed in young star clusters in starburst and interacting galaxies, which seems to be close to
20 \% (Meurer et al.\ 1995, Zepf et al.\ 1999). Some of these young clusters might however be unbound,
or become unbound due to the mass-loss of individual cluster stars. This could bring both values
into agreement. 

\section{Acknowledgements} 
I thank Douglas Heggie for a careful reading of the paper and comments. 
H.B.\ is supported by PPARC under grant 1998/00044.


\begin{references}
\reference Baumgardt, H. 2001, MNRAS, in press, astro-ph/00120330
\reference Binney, J., Tremaine, S. 1987, Galactic Dynamics, Princeton Univ.\ Press, Princeton
\reference Fukushige, T., Heggie, D.\ C. 2000, MNRAS,318, 753  
\reference Giersz, M., Heggie, D.\ C. 1996, MNRAS, 279, 1037
\reference Grillmair, C.\ J., Freeman, K.\ C., Irwin, M., Quinn, P.\ J. 1995, AJ, 109, 2553
\reference Harris, W.\ E. 1996, AJ, 112, 1487
\reference Harris, W.\ E. 2000, in Globular Cluster Systems: Lectures for the 1998 Saas-Fee Advanced
   Course on Star Clusters, in press
\reference Heggie, D.\ C.\ et al. 1998, in
  Highlights of Astronomy, Vol.\ 11B, Ed.\ J.\ Andersen, D.\ Reidel Publ.\ Comp.\
   Dordrecht, p.\ 591  
\reference Helmi, A., White, S.\ D.\ M., de Zeeuw, P.\ T., Zhao, H. 1999, Nature, 402, 53
\reference Kundu, A.\ et al. 1999, ApJ, 513, 733
\reference Leon, S., Meylan, G., Combes, F. 2000, A\&A, 359, 907
\reference Majewski, S.\ R., Munn, J.\ A., Hawley, S.\ L. 1996, ApJ, 459, L73
\reference Meurer, G.\ R.\ et al. 1995, AJ, 110, 2665
\reference Suntzeff, N.B., Kinman, T.D., Kraft, R.P. 1991, ApJ, 367, 528 
\reference Whitmore, B.\ C., Schweizer, F. 1995, AJ, 109, 960
\reference Whitmore, B.\ C.\ et al. 1999, AJ, 118, 1551
\reference Zepf, S.\ E.\ et al. 1999, AJ, 118, 752
\end{references}
\end{document}